\documentclass[prl,aps,amssymb,showpacs,twocolumn]{revtex4-1}
\usepackage{amsmath}
\usepackage{amssymb}
\usepackage{amsthm}
\usepackage{amsfonts}
\usepackage{enumerate}
\usepackage{latexsym}
\usepackage{color}
\usepackage{bm}
\usepackage{graphicx}
\usepackage{subfigure}

\newcommand{\beq}{\begin{equation}}
\newcommand{\eneq}{\end{equation}}

\newcommand{\bs}[1]{\boldsymbol{#1}}

\input{epsf}

\begin{document}

\tolerance 10000

\newcommand{\vk}{{\bf k}}


\title{Mechanism for a Pairing State with Time-Reversal Symmetry Breaking in Iron-Based Superconductors}

\author{Christian Platt${}^1$}
\author{Ronny Thomale${}^{1,2,3}$}
\author{Carsten Honerkamp${}^4$}
\author{Shou-Cheng Zhang${}^3$}
\author{Werner Hanke${}^1$}

\affiliation{${}^1$Theoretical Physics, University of W\"urzburg, D-97074 W\"urzburg}
\affiliation{${}^2$Department of Physics, Princeton University, Princeton, NJ 08544}
\affiliation{${}^3$Department of Physics, McCullough Building, Stanford University, Stanford, California 94305-4045}
\affiliation{${}^4$Institute for Theoretical Solid State Physics, RWTH Aachen University and \\JARA - FIT Fundamentals of Future Information Technology, D-52056 Aachen}

\begin{abstract}
  The multipocket Fermi surfaces of iron-based superconductors 
  promote pairing states with both $s_{\pm}$-wave and
  $d_{x^2-y^2}$-wave symmetry.  We argue that the competition between
  these two order parameters could lead to a time-reversal-symmetry
  breaking state with $s+id$-pairing symmetry in the iron-based superconductors, and propose
  serveral scenarios in which this phase may be found.  To understand
  the emergence of such a pairing state on a more rigorous footing, we
  start from a microscopic 5-orbital description representative for
  the pnictides.  Using a combined approach of functional
  renormalization group and mean-field analysis, we identify the
  microscopic parameters of the $s+id$-pairing state.
  There, we find the most promising region for $s+id$-pairing in the electron doped
  regime with an enhanced pnictogen height. 
\end{abstract}

\date{\today}

\pacs{74.20.Mn, 74.20.Rp, 74.25.Jb, 74.72.Jb}

\maketitle

Iron based superconductors (SC) offer an appealing platform to investigate
the interplay among pairing interactions, pairing symmetries and Fermi
surface topologies~\cite{kamihara-08jacs3296}. Generally, repulsive
interactions in momentum space can lead to a change of sign in the
pairing amplitude. A large class of iron based SC have
disconnected Fermi surface pockets, consisting of hole pockets at the
$\Gamma=(0,0)$ and possibly $M=(\pi,\pi)$ points, and two electron pockets at the $X=(\pi,0)/(0,\pi)$ points in the
unfolded Brillouin zone (BZ) with one iron atom per unit cell. When the
repulsive interactions between the hole and the electron pockets
dominate, an $s_\pm$ pairing symmetry can be
obtained~\cite{mazin-08prl057003,kuroki-08prl087004,chubukov-08prb134512}. On
the other hand, when the repulsive interactions between the two
electron pockets dominate, a propensity toward $d$-wave pairing
symmetry can be expected. When both types of interactions are
comparable, there is hence a frustration between the two pairing
symmetry types. A recent theoretical proposal suggests that the system
can resolve the frustration by a pairing state with the $s+id$ pairing
symmetry which spontaneously breaks time-reversal (TR)
symmetry~\cite{Lee:2009}.  The possibility of a TR-symmetry breaking
pairing state due to frustrating pairing
interactions among three or more Fermi pockets has also been
investigated in several other
contexts~\cite{Agterberg:1999,Ng:2009,Stanev:2010,Tanaka:2010,Hu:2011}.
In general, time reversal breaking pairing states have rather
accessible experimental signatures, and several proposals have been
suggested in the context of iron based
SC~\cite{Lee:2009}.

\begin{figure}[t!]
  \begin{minipage}[l]{0.99\linewidth}
    \includegraphics[width=\linewidth]{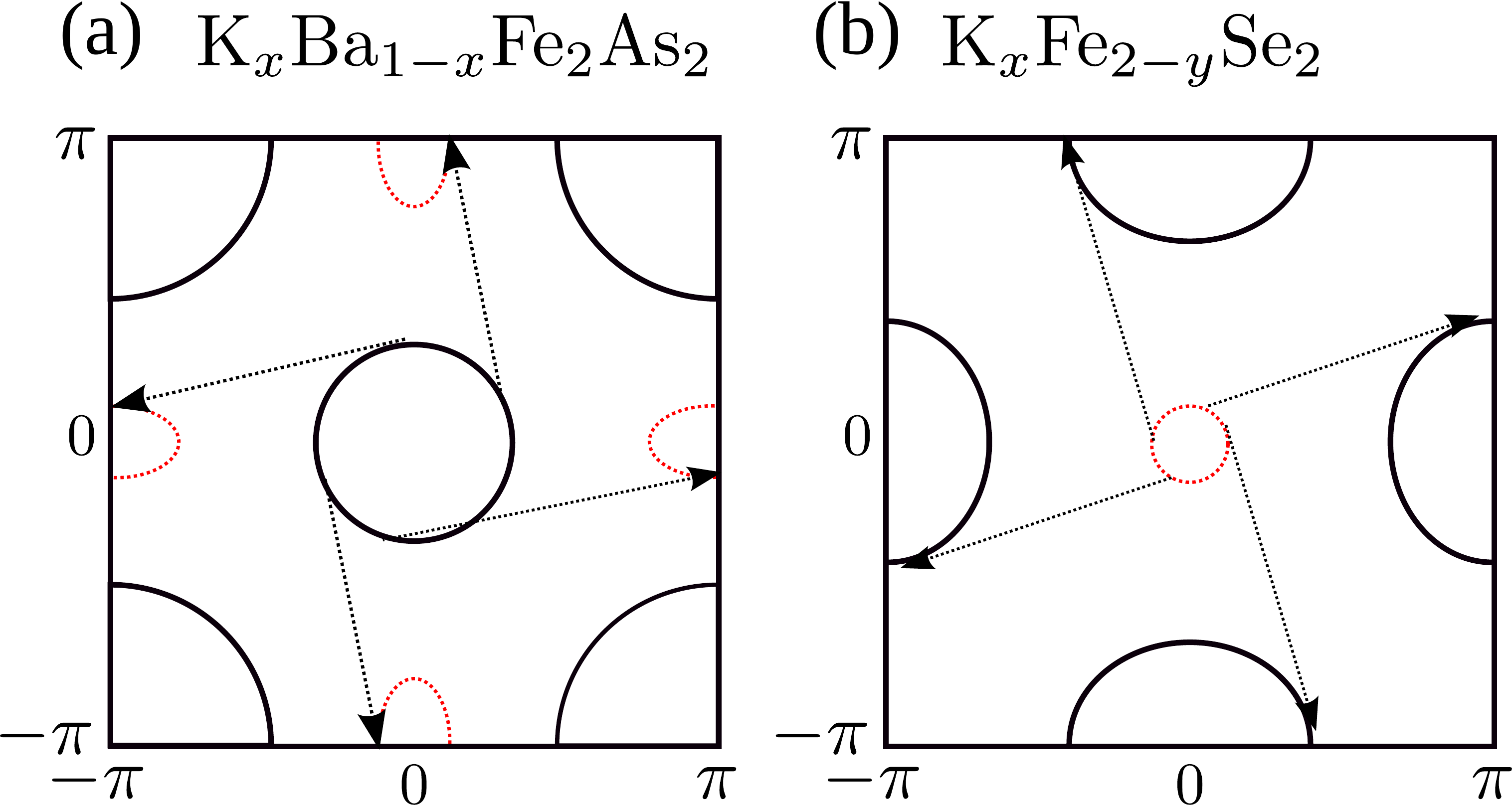}
  \end{minipage}
  \caption{(Color online). Frustrating the d-wave limits of
     $\text{K}\text{Fe}_2\text{As}_2$ (a) and $\text{K}_x\text{Fe}_{2-y}Se_{2}$ (b). Upon doping or differently
    induced band structure effects, electron pockets appear (dashed red) in (a) and
    a hole pocket appears (dashed red) in (b) which populate the $q \sim (\pi,0) /
    (0,\pi)$ scattering channels enhancing the $s_\pm$ symmetry. This
    leads to frustration providing the background for $s+id$ pairing.}
\label{pic0}
    \vspace{-0pt}
\end{figure}

In principle there are various experimentally tunable parameters to
drive the competition between $s$-wave and $d$-wave in the pnictides,
giving the opportunity to start from both limits.  In
$\text{K}_x\text{Ba}_{1-x}\text{Fe}_2\text{As}_2$, the Fermi surface
topology can be chosen as a paradigmatic setup for $s_\pm$, consisting of hole
pockets at $\Gamma$ and the electron pockets at $X$
for optimal doping $x\simeq 0.4$. For
increasing $x$, however, the electron pockets decrease, and have
nearly disappeared for $x=1$ [Fig.~\ref{pic0}], which has been
recently suggested to host a $d$-wave pairing
symmetry~\cite{Thomale:2010}.  In this system, it is hence plausible
that a $s+id$ pairing state can be realized for intermediate
values of $x$. In the chalcogenide $\text{K}_x\text{Fe}_{2-y}Se_{2}$,
the electron pockets at the $X$ points dominate, and, for a situation
seemingly inverse to $\text{K}\text{Fe}_2\text{As}_2$, a $d$-wave
pairing symmetry may likewise be expected~\cite{hirschi:2011,fango:2011}. (It
should be noted that the actual pairing symmetry in the
chalcogenides is far from settled, as a strong
coupling perspective may likewise suggest $s_\pm$
pairing~\cite{Fang:2011}.) By tuning doping or other possible
parameters affecting the band structure such as pressure, one possibly induces a
pocket at $\Gamma$, increasing the tendency towards $s_\pm$ pairing
symmetry [Fig.~\ref{pic0}]. In this case, one could also expect an
$s+id$ pairing state.  By
systematically tuning the Fermi pocket topologies, one can compare the
predicted pairing symmetries with experiments, and determine the
nature of the pairing interaction by these investigations, starting
from compound settings with a suspected $d$-wave symmetry
[Fig.~\ref{pic0}].

In the following, we rather intend to start from an $s_\pm$ pairing
state to begin with, and address how we can enhance the
competitiveness of the $d$-wave symmetry to drive the system into the
$s+id$ regime. The reason for this is two-fold. First, the $s_\pm$
symmetry is much more generic for the different classes of
pnictides. Second, as we will see below, we find the most promising
setup to be located on the electron doped side of pnictides, where
high-quality samples have already been grown for different
families. We hence believe that this regime may be the experimentally
most accessible scenario at the present stage, which is why we
explicate it in detail.  In this paper, we investigate the microscopic
mechanism of the $s+id$ pairing state by the functional
renormalization group (fRG) method of a five band model. We systematically
vary the doping level and the strength of intra-orbital interaction,
which determine the ratio between the electron-hole pocket and the
electron-electron pocket mediated pairing interactions. 
In this microscopic investigation, we
find that the $s+id$ pairing state can be realized in the
intermediate electron-doped regime, given that we also adjust the
pnictogen height parameter of the system appropriately.

\begin{widetext}
 \begin{figure*}[t!]
  \begin{minipage}[l]{0.99\linewidth}
    \includegraphics[width=\linewidth]{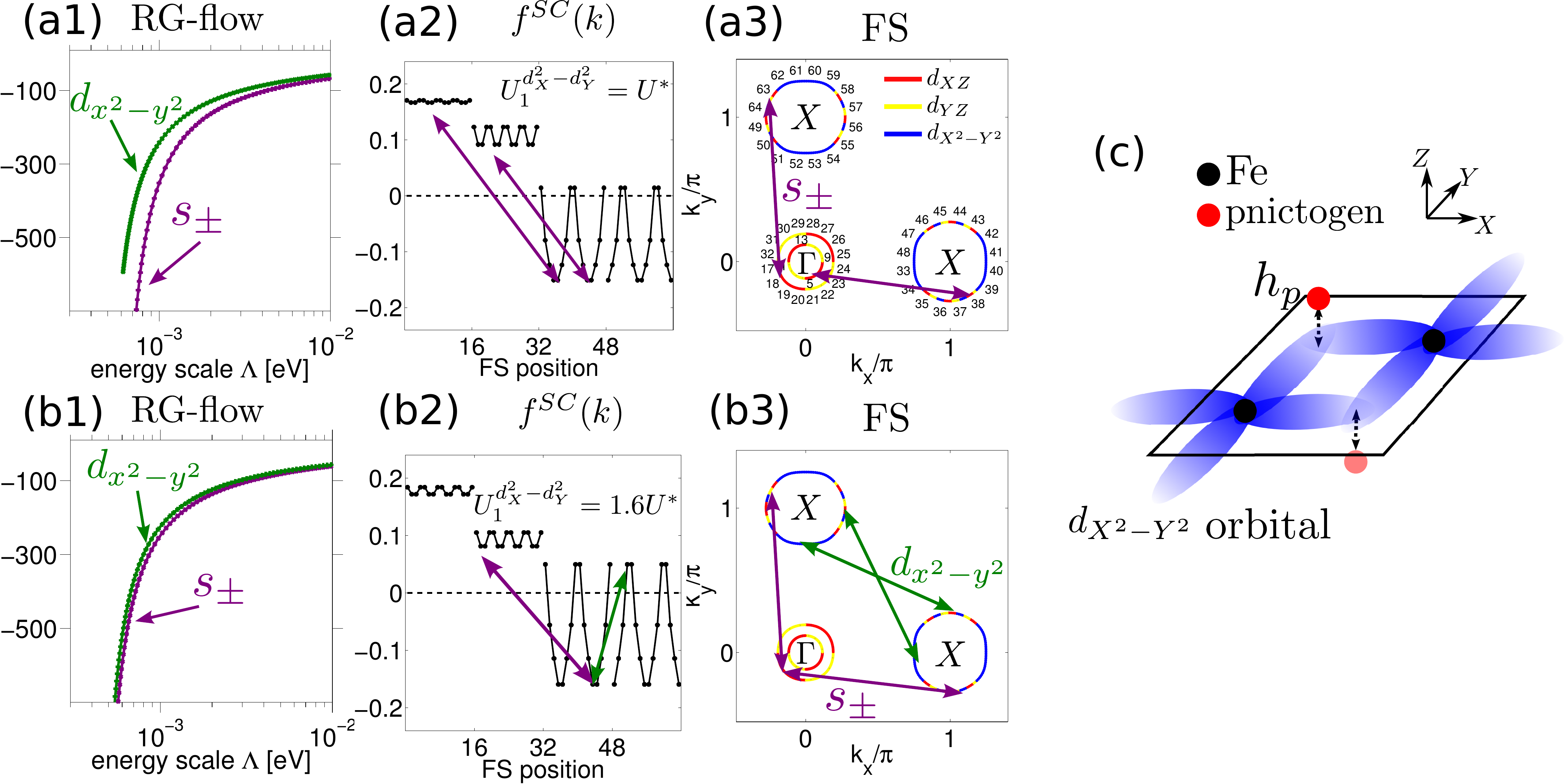}
  \end{minipage}
  \caption{(Color online). Competing pairing orders and SC
    form factors for $U_1(d_{X^2-Y^2})=U_1^*=2.5eV$ (a) and
    $U_1(d_{X^2-Y^2})=1.6U^*$ (b) at electron doped filling $n=6.13$. 
    RG channel flow (a1,b1) and $s_{\pm}$-gap
    form factor (a2,b2). $s_{\pm}/d_{x^2-y^2}$
    transition from (a) to (b): increasing $U_1(d_{X^2-Y^2})$ enhances the
    gap anisotropy of the $s_{\pm}$-form factor on the
    electron pockets [${\bs k}$-patching: points 33-64 see (a3)] shown in
    (a2,b2) until its symmetry switches to
    $d_{x^2-y^2}$. (a3,b3) Interactions mediated by $U_1$, setting up the $s_{\pm}$-pairing
    tendency between hole pockets and electron pockets $(\Gamma\leftrightarrow
    X)$ and the competing $d_{x^2-y^2}$-pairing symmetry due to electron-electron
    $(X\leftrightarrow X)$ interaction. (c) The variation of the
    pnictogen height $h_p$ particularly affects the spread of the $d_{X^2-Y^2}$-orbital and therefore $U_1(d_{X^2-Y^2})$, as it is oriented to the planar projection of the pnictogen.}
\label{pic1}
    \vspace{-0pt}
 \end{figure*}
\end{widetext}

\begin{figure}[t!]
  \begin{minipage}[l]{0.99\linewidth}
    \includegraphics[width=\linewidth]{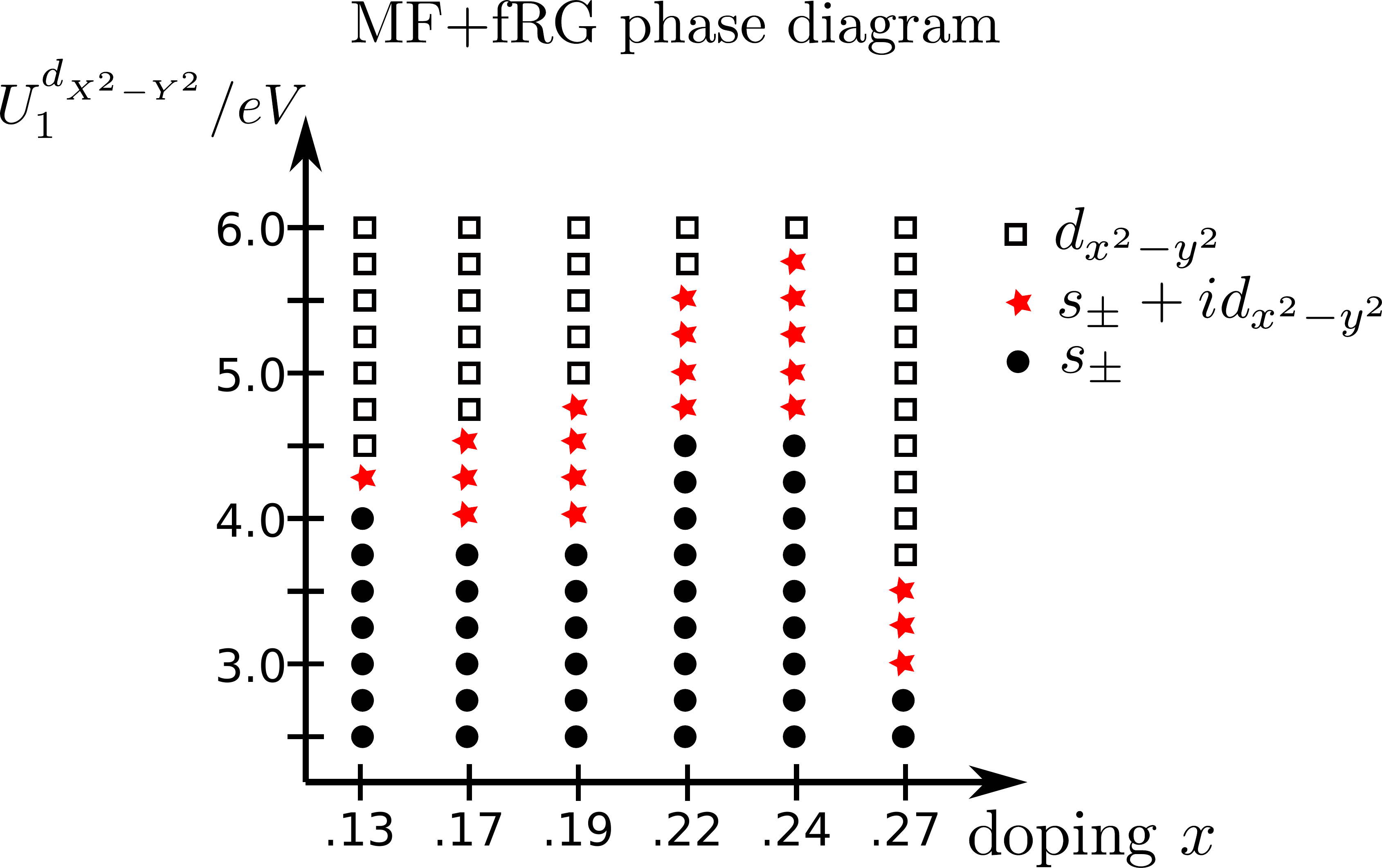}
  \end{minipage}
  \caption{(Color online).\label{pic2} Preferred pairing as a function
    of electron doping and intra-orbital Coulomb interaction
    $U_1(d_{X^2-Y^2})$. The results are
    obtained by minimizing the mean-field free energy of the effective
    theory taken from fRG at $\Lambda\approx .001eV$. At $27\%$
    electron doping, the $s+id$-pairing state occurs at
    $U_1(d_{X^2-Y^2})=3eV$ which is comparable to the intra-orbital
    repulsion in the remaining orbitals $U_1 = 2.5eV$.}
    \vspace{-0pt}
\end{figure}

We start from a representative 5-band model for the pnictides which is
obtained from LDA-type calculations~\cite{kuroki-08prl087004}. It has
been considered by us before as a starting point for explaining the
difference between the isovalent P-based and As-based
pnictides~\cite{Thomale:2011}. The LDA ''non-interacting'' part is
given by
\begin{equation}\label{h0}
H_0 = \sum_{\bs{k},s} \sum_{a,b=1}^{5}c_{\bs{k}as}^{\dagger}K_{ab}(\bs{k})c_{\bs{k}bs}^{\phantom{\dagger}}.
\end{equation}
Here $c'$s stand for electron annihilation operators, $a,b$ for the
$d$-orbitals, $s$ denotes the spin indices and $K_{ab}(\bs{k})$ the
orbital (i.e. maximally-localized Wannier function) matrix elements of
the Kohn-Sham Hamiltonian. The band structure features electron
pockets at $X$ and hole pockets at $\Gamma$, which is the typical
situation in the pnictides [Fig.~\ref{pic1}] for sufficient electron
doping.  The many-body interaction part is given by the intra- and
inter-orbital interactions $U_1$ and $U_2$, as well as the Hund's
coupling $J_{\text{H}}$ and the pair hopping $J_{\text{pair}}$:
\begin{eqnarray}
\label{hint}
&&H_{\text{int}}=\sum_i \left[ U_1 \sum_{a} n_{i,a\uparrow}n_{i,a\downarrow} + U_2\sum_{a<b,s,s'} n_{i,as}n_{i,bs'} \right.\nonumber \\
&& \hspace{-15pt}\left.+\sum_{a<b}(J_{\text{H}} \sum_{s,s'} c_{ias}^{\dagger}c_{ibs'}^{\dagger}c_{ias'}^{\phantom{\dagger}}c_{ibs}^{\phantom{\dagger}}  +J_{\text{pair}} c_{ia\uparrow}^{\dagger}c_{ia\downarrow}^{\dagger}c_{ib\downarrow}^{\phantom{\dagger}}c_{ib\uparrow}^{\phantom{\dagger}}) \right]\hspace{-4pt},
\end{eqnarray}
where $n_{i,as}$ denote density operators at site $i$ of spin $s$ in
orbital $a$.  Typical interaction settings are dominated by
intra-orbital coupling, $U_1 > U_2 > J_{\text{H}} \sim
J_{\text{pair}}$.
In the
fRG~\cite{Wang:2009,Platt:2009,Thomale:2009,Thomale:2011,Honerkamp:2001,Metzner:2011},
one starts from the bare many-body interaction~\eqref{hint} in
the Hamiltonian. The pairing is dynamically generated by
systematically integrating out the high-energy degrees of freedom
including important fluctuations (magnetic, SC, screening, vertex
corrections) on equal footing. This differs from the RPA which takes
right from the outset a magnetically driven SF-type of pairing
interaction.  For a given instability characterized by some order
parameter $\hat{O}_{{\bf k}}$, the effective interaction vertex
$V_{\Lambda}({\bf k}_1,{\bf k}_2,{\bf k}_3,{\bf k}_4)$ in the
particular ordering channel can be written in shorthand notation as
$\sum_{{\bf k},{\bf p}}V_{\Lambda}({\bf k},{\bf p})
[\hat{O}^{\dagger}_{{\bf k}} \hat{O}^{\phantom{\dagger}}_{{\bf
    p}}]$.  Accordingly, the effective interaction vertex $V_{\Lambda}({\bf
  k},{\bf -k},{\bf p},{\bf -p}))$ in the Cooper channel can be
decomposed into different eigenmode
contributions~\cite{Wang:2009,Thomale:2011}
\begin{equation}\label{decomp}
V^{\text{\text{SC}}}_{\Lambda} ({\bf k},{\bf p})= \sum_i c_i^{\text{\text{SC}}}(\Lambda) f^{\text{\text{SC}},i}({\bf k})^* f^{\text{\text{SC}},i}({\bf p}),
\end{equation}
where $i$ is a symmetry decomposition index,
and the leading instability of that channel corresponds to an
eigenvalue $c_1^{\text{\text{SC}}}(\Lambda)$ first diverging under the
flow of $\Lambda$. $f^{\text{\text{SC}},i}(\bf{k})$ is the SC form
factor of pairing mode $i$ which tells us about the SC pairing
symmetry and hence gap structure associated with it. In the fRG, from
the final Cooper channel in the effective interaction vertex, 
this quantity is computed along
the discretized Fermi surfaces [Fig.~\ref{pic1}(a3)], and the leading
SC instabilities are plotted in Figs.~\ref{pic1}(a1) and (b1). The
interaction parameters are kept fixed at the representative setup $U_1
= 2.5 eV, U_2 = 2.2 eV,J_{\text{H}}=1.2 eV, J_{\text{pair}}=0.2 eV$
($U_1$ for the $d_{X^2-Y^2}$-orbital is varied as explicitly stated in
Figs.~\ref{pic1} and \ref{pic2}).  The relatively large bare value of $J_H$ is motivated partly by recent findings, in
particular, for a sizable Hund's rule coupling~\cite{Zhang:2009,
  Haule:2009}.  Furthermore, as a parameter trend, larger $J_H$ and
smaller $J_{\text{pair}}$ tends to prefer the $s+id$-phase in the
electron-doped regime for rather moderate values of intra-orbital
coupling $U_1$ [Fig.~\ref{pic2}].

The situation in Fig.~\ref{pic1} is representative for moderate
electron doping and interaction scales of the pnictides, where the
$\Gamma \leftrightarrow X$ pair scattering between the hole pockets at
$\Gamma$ and the electron pockets at $X$
dominates. Already from the BCS gap equation, a finite momentum
transfer can induce pairing only when the wave vector of such an
interaction connects regions on one FS (such as in the cuprate case),
or regions on different FSs (such as in the pnictide case), which have
opposite signs of the SC order parameter. This corresponds to putting
the electron pairs in an anisotropic wave function such as
sign-reversing $s$-wave ($s_{\pm}$) in Fig.~\ref{pic1}a, where the wave
vector ($\pi,0$) in the unfolded BZ connects hole and
electron pockets with a sign-changing $s_{\pm}$
gap~\cite{mazin-08prl057003,chubukov-08prb134512}.
However, in the fRG calculation of Fig.~\ref{pic1}b with increased
$U_1$ interaction on the $d_{X^2-Y^2}$ orbital, a green arrow for
$X\leftrightarrow X$ scattering indicates additional interactions that
become similarly important as the $(\pi,0)$ channel. This increased
$U_1$ can be tuned by the pnictogen height as explained below
and frustrates the previous ''pure'' $s_{\pm}$ limit
($\Gamma\rightarrow X$). The system then strikes a
compromise~\cite{Platt:2010,Thomale:2011} by enhancing the anisotropy
of the formfactor (denoted by $f^{\text{SC}}({\bf k})$ in
Fig.~\ref{pic1}) on the electron pockets at $X$.
The multi-band SC hence adjusts the momentum dependence of the gap,
i.e.  its anisotropy, so as to minimize the effect of the Coulomb
repulsion which is set up by the competition between $s_{\pm}$ and
$d_{x^2-y^2}$-wave channels.

We now have all ingredients to tune the pairing symmetry from
$s_{\pm}$-wave to $d_{x^2-y^2}$-wave and, eventually, into the
TR-symmetry broken $s+id$ phase.  In most of the iron-based SC,
the tendency towards $s_{\pm}$-pairing occurs slightly more pronounced
than the competing $d_{x^2-y^2}$-pairing and, at first glance, the
resulting frustration appears to be too small for causing
$s+id$-pairing.  Therefore, in order to increase frustration, we
somehow have to enhance the pair-scattering between the electron
pockets at $X$ which then promotes the sub-leading
$d_{x^2-y^2}$-channel.  As shown in a-priori determinations of the
interaction in Eq.~(\ref{hint}), expressed in terms of orbital matrix elements,
the pnictogen height~$h_p$ (measured from the Fe-plane [Fig.~\ref{pic1}c]) has a
substantial influence on the intra-orbital interaction $U_1$ between
$d_{X^2-Y^2}$-Wannier orbitals~\cite{Imada:2010}, which can be either
modified by isovalent doping or pressure. By increasing $h_p$, the
Wannier functions in this orbital are further localized, causing an
increase of $U_1(d_{X^2-Y^2})$. In Fig.~\ref{pic1}b, we have already
used this fact to demonstrate that, for moderate e-doping (13\%),
large values of this matrix element drive the SC instability from
$s_{\pm}-$ to $d_{x^2-y^2}$-wave symmetry.

For this setup, we present our predictions for TR-symmetry breaking in
a schematic phase diagram
in Fig.~\ref{pic2}, where we plot the leading $s_{\pm}$, $d_{x^2-y^2}$
and finally $s+id$ SC solutions as a function of $U_1(d_{X^2-Y^2})$,
and electron doping.
There, we have used our fRG result as a starting point for a renormalized mean field analysis~\cite{Reiss:2007}.
In this MF+fRG approach, the one-loop flow is stopped at a scale
$\Lambda$ which is small compared to the bandwidth, but still safely
above the scale $\Lambda_c$, where the 2-particle vertex diverges. In
this range, the particular choice of the cutoff $\Lambda$ does not
significantly influence the results in Fig.~\ref{pic2}. The
renormalized coupling function
$V^{\Lambda}(\bs{k}_1,\bs{k}_2,\bs{k}_3,\bs{k}_4)$ is taken as an
input for the mean field treatment of the remaining modes.  As shown
in Fig.~\ref{pic1}, the regime of $s_{\pm}/d$-wave pairing competition
features a single channel SC instability without other competing
(e.g. magnetic) instabilities and, therefore, justifies
\begin{equation}\label{meandecomp}
V^{\Lambda}(\bs{k}_1, \bs{k}_2, \bs{k}_3, \bs{k}_4) \approx V^{pair}(\bs{k}_1,\bs{k}_3)\delta_{\bs{k}_2,-\bs{k}_1}\delta_{\bs{k}_4,-\bs{k}_3},
\end{equation}
with
$V^{pair}(\bs{k}_1,\bs{k}_3)=V^{\Lambda}(\bs{k}_1,-\bs{k}_1,\bs{k}_3,-\bs{k}_3)$.
The effective theory for quasiparticles near the Fermi surface
$(|\xi(\bs{k})|<\Lambda)$ is modeled by the reduced Hamiltonian
\begin{equation}\label{red}
H^{\Lambda} = \sum_{\bs{k}s}\xi(\bs{k})c_{\bs{k}s}^{\dagger}c_{\bs{k}s}^{\phantom{\dagger}} + \frac{1}{N}\sum_{\bs{k},\bs{q}}V^{pair}(\bs{k},\bs{q})c_{\bs{k}\uparrow}^{\dagger}c_{-\bs{k}\downarrow}^{\dagger}c_{-\bs{q}\downarrow}^{\phantom{\dagger}}c_{\bs{q}\uparrow}^{\phantom{\dagger}},
\end{equation}
where $\xi(\bs{k})$ is taken as the bare dispersion due to only weak
band-renormalization effects.  The MF solution of this reduced
Hamiltonian is obtained as in BCS theory, by solving the self-consistent gap-equation
and calculating the corresponding grand potential
which is
\begin{equation}\label{estimate}
\Omega^{stat}=-\sum_{\bs{k}} \frac{|\Delta_{\bs{k}}|^2 + 2\xi(\bs{k})^2}{2\sqrt{\xi(\bs{k})^2 + |\Delta_{\bs{k}}|^2}} + \sum_{\bs{k}}\xi(\bs{k}).
\end{equation}
Within a reasonable range of parameters for the electron-doped pnictides, we then find a regime favoring $s+id$-pairing due to
\begin{equation}
\Omega^{stat}_{s+id}<\Omega^{stat}_{s_{\pm}},\Omega^{stat}_{d}.
\end{equation}
 
Note, that this phase regime is only a lower bound for the existence of $s+id$ which
probably is much larger. This is because the fRG setup at present only
allows us to obtain the leading SC instability at some finite
$\Lambda_c$, while the $s+id$ phase may well set in below
$\Lambda_c$. This would manifest itself as a change of the SC phase as
a function of temperature.

In summary, we have presented a microscopic analysis, based on
a-priori electronic structure determinations and a combination of the
fRG with an MF treatment of the remaining low-energy states,
to derive a kind of "guiding principle'' for a possible $s+id$
pairing state in the pnictides. 
For the case of increased electron
doping and pnictogen height, we have illustrated how this drives
the system into an $s+id$ SC state. Aside from this example, other
regimes in the pnictides likewise promise the possible realization of
an $s+id$ state, such as hole-doped (K,Ba)-122 interpolating between
the $s$-wave limit $(x\sim 0.4)$ and $d$-wave limit $(x\sim 1)$ as
well as possibly the chalcogenides
$\text{K}_x\text{Fe}_{2-y}Se_{2}$.

\begin{acknowledgments}
  We thank A. V. Chubukov, D. Scalapino, and P. Hirschfeld for discussions.  CP, RT, CH and WH are supported by
  DFG-SPP 1458/1, CP, CH and WH by DFG-FOR 538. RT is
  supported by the Humboldt Foundation.
\end{acknowledgments}



\end{document}